\documentstyle[12pt]{article}

\newcommand{\tr}{{\rm tr}}
\newcommand{\Tr}{{\rm Tr}}

\title{Bosonization and Duality \\ of \\ Massive Thirring Model}
\author{%
  Kei-Ichi Kondo
  \thanks{e-mail: kondo@cuphd.nd.chiba-u.ac.jp}
  \vspace{1.2em}
  \\
  Department of Physics, Faculty of Science \\
  $\&$ Graduate School of Science and Technology,\\
         Chiba University, Chiba 263, Japan
  \thanks{Permanent  address.
  Address from September 1995  to February 1996:
  HLRZ c/o KFA J\"ulich, D-52425 J\"ulich, Germany,
  and
  address from March 1996  to December 1996:
  Department of Physics, Theoretical Physics,
  University of Oxford,
  1 Keble Road, Oxford, OX1 3NP, UK.}
}
\date{CHIBA-EP-88-REV,
\\ February 1995,
\\ hep-th/9502100}
\begin{document}
\maketitle
\centerline{Abstract}
 Starting from a reformulation of the Thirring model as a
gauge theory, we consider the bosonization of the
$D$-dimensional multiflavor massive Thirring model $(D \ge
2)$ with four-fermion interaction of the current-current
type.  Our method leads to a novel interpolating Lagrangian
written in terms of two gauge fields. Especially we pay
attention to the case of very massive fermion $m \gg 1$ in
(2+1) and (1+1) dimensions.   Up to the next-to-leading
order of $1/m$, we show that the (2+1)-dimensional massive
Thirring model is mapped to the Maxwell-Chern-Simons theory
and that the (1+1)-dimensional massive Thirring model is
equivalent to the massive free scalar field theory. In the
process of the bosonization of the Thirring model, we point
out the importance of the gauge-invariant formulation.
Finally we discuss a possibility of extending this method
to the non-Abelian case.

\newpage
\section{Introduction and main results}
\setcounter{equation}{0}
\par
Recently the Thirring model \cite{Thirring}
was reformulated as a gauge theory \cite{IKSY95} and was
identified with a gauge-fixed version of the corresponding
gauge theory by introducing the St\"uckelberg field
$\theta$ in addition to the auxiliary vector field
$A_\mu$.  In this formulation, the auxiliary field $A_\mu$
is identified with the gauge field.
In the previous paper \cite{Kondo95a}, we have given
another reformulation of the Thirring model as a gauge
theory based on the general formalism for the
constrained system, so called the Batalin-Fradkin-Vilkovisky
(BFV) formalism \cite{BF86}.  Especially Batalin-Fradkin
(BF) method \cite{BF87} gives the general procedure by
which the system with the second class constraint is
converted to that with the first class one.  The new field
which is necessary to complete this procedure is called the
Batalin-Fradkin field
\cite{BF87}.  In the massive gauge theory the
Batalin-Fradkin field is nothing but the well-known
St\"uckelberg field as shown in
\cite{FIK90}.
\par
We consider the mapping from quantum field theory of
interacting fermions onto an equivalent theory of
interacting bosons.
In this paper such an equivalent bosonic theory
to the original {\it massive} Thirring model is obtained
starting from the formulation of the
$D$-dimensional Thirring model $(D=d+1 \ge 2)$ as a gauge
theory.  This is a kind of bosonization.
The bosonization of the (1+1) dimensional Thirring model
has been studied by many authors and is well known, see
for example
\cite{Klaiber68,Coleman75,Halpern75,Witten84,GSS81,PW83,VDP84,Dorn86,Rothe86,DV88}.
In this paper we consider the bosonization of the
multiflavor massive Thirring model in $D=d+1 \ge 2$
dimensions. Especially, we study the large fermion
mass limit $m \gg 1$, in (2+1) and (1+1) dimensional cases
explicitly. A motivation of studying the multiflavor case
stems from the renormalizability of the Thirring model in
$1/N$ expansion at least for $D=3$ \cite{RWP91}, although it
is perturbatively non-renormalizable for $D>2$.
\par
In the case of (2+1)-dimensions, we show that,
up to the next-to-leading order in the inverse fermion
mass, $1/m$, {\it the (2+1)-dimensional massive Thirring
model is equivalent to the Maxwell-Chern-Simons theory, the
topologically massive $U(1)$ gauge theory}.  This
equivalence in three dimensions has been shown, to the
lowest order in $1/m$ for a massive Thirring
model, by  Fradkin and Schaposnik \cite{FS94}.
The fermi-bose equivalence was discussed earlier also in
\cite{DR88,DM92}.
In the multiflavor case we consider in this paper,
there are many possibilities of taking the massive limits.
Indeed we point out that there is a possibility that the
leading order term, i.e., Chern-Simons term vanishes and
hence only the non-local Maxwell-like term remains in the
next-to-leading order, depending on the configuration of
fermion masses.  In such a case we are forced to consider
the next-to-leading order of $1/m$.
As discussed in the previous paper \cite{Kondo95a}, the
existence of the kinetic term for the gauge field $A_\mu$
is essential in formulating the Thirring model as a gauge
theory based on the BFV formalism.
Such a configuration of fermion masses has important
implications from the viewpoint of chiral symmetry breaking
for 4-component fermions\cite{Kondo95a}.
This is in sharp contrast to the case treated in
\cite{FS94} where the non-vanishing Chern-Simons term is
assumed from the beginning and hence only the leading
order term is taken into account for bosonization.
In our bosonization scenario the gauge invariance should be
preserved in the step of calculations. In this connection we
mention the choice of appropriate regularization.
\par
The method of Fradkin and Schaposnik is more elegant than
ours.  However we can raise the following questions on their
treatment of bosonization.
\begin{enumerate}
\item
The interpolating Lagrangian ${\cal L}_{FS}$ of Deser and
Jackiw \cite{DJ84} was introduced as a device for showing
this equivalence in \cite{FS94}:
\begin{eqnarray}
 {\cal L}_{FS}[V_\mu,H_\mu] =  {1 \over 2} V^\mu V_\mu
 - {1 \over 2}\epsilon^{\mu\nu\rho} V_\rho
 (\partial_\mu H_\nu - \partial_\nu H_\mu)
 + {2\pi \over G}\epsilon^{\mu\nu\rho} H_\mu \partial_\nu
H_\rho.
\end{eqnarray}
However the origin of the
interpolating Lagrangian was never shown.  How can we
derive this type of interpolating Lagrangian from the
Thirring model or the Maxwell-Chern-Simons theory?
\item
 The original Thirring model has no gauge symmetry.
Nevertheless the equivalent bosonic theory, the
Maxwell-Chern-Simons theory, has $U(1)$ gauge symmetry.
Where does this gauge symmetry come from?
\item
 The interpolating Lagrangian is invariant under a gauge
transformations for the gauge field $H_\mu$, while the
auxiliary vector field $V_\mu$ does not have the gauge
symmetry: ${\cal L}_{FS}$ is invariant under
$\delta H_\mu = \partial_\mu \lambda$ and
$\delta V_\mu = 0$.
 In integrating out the the gauge field $H_\mu$ in the
interpolating Lagrangian, they had to introduce the
gauge-fixing term ad hoc.
\end{enumerate}
In contrast to \cite{FS94}, we start from the
gauge-invariant or more precisely the BRS-invariant
formulation of the Thirring model and discuss the
bosonization of the massive Thirring model.
The bosonization of the massless Thirring model was already
discussed based on this strategy in \cite{IKSY95}.
However the bosonization in the usual or exact sense is
possible only in the massive case except for $D=2$, as
shown in this paper. Our approach is more direct than the
method of \cite{FS94} and is able to {\it derive} a novel
interpolating Lagrangian as a natural consequence (in an
intermediate step) of bosonization.  Our interpolating
Lagrangian is written in terms of two gauge fields which
have independent gauge symmetries.  Thanks to the gauge
invariant formulation, we can fix definitely the gauge
invariance appearing in the interpolating Lagrangian
\cite{DJ84}.
It is interesting to extend our method into the
non-Abelian case \cite{DJ84,HH92}.  This issue will be
discussed in the final section.
\par
The (1+1)-dimensional case is also discussed, although the
gauge symmetry disappears in this case. The Thirring model
in (1+1)-dimensions is rewritten into the equivalent scalar
field theory
\cite{Klaiber68,Coleman75,Halpern75}.  It is well known that
the (1+1)-dimensional {\it massless} Thirring model is
exactly solvable in the sense that the model is equivalent
to the massless {\it free} scalar theory
\cite{Thirring}.   In this paper we show, as a special
case of the above formalism, {\it the massive Thirring
model in (1+1)-dimensions is equivalent to the massive
free scalar field theory in (1+1)-dimensions}, up to the
next-to-lowest order in $1/m$.
This is consistent with the well-known result.
\par
In the previous paper \cite{Kondo95a} we have studied the
spontaneous breakdown of the chiral symmetry in the
massless Thirring model, $m \rightarrow 0$. In this paper we
consider another extreme limit $m \rightarrow \infty$.
According to the universality hypothesis, the critical
behavior of the model will be characterized by a small
number of parameters appearing in the original Lagrangian
of the model such as symmetry, range of interaction and
dimensionality.
Therefore, in the large fermion mass limit, the
critical behavior of the Thirring model will be
characterized by studying the equivalent bosonic theory
according to the above bosonization.

\section{Bosonization}
\setcounter{equation}{0}
The Lagrangian of the $D$-dimensional multiflavor Thirring
model  ($D=d+1 \ge 2$) is given by
\begin{eqnarray}
 {\cal L}_{Th}
 &=& \bar \psi^j i \gamma^\mu \partial_\mu \psi^j
 - m_j \bar \psi^j \psi^j
- {G \over 2N}
(\bar \psi^j  \gamma_\mu \psi^j)
(\bar \psi^k  \gamma^\mu \psi^k),
\label{th}
\end{eqnarray}
where $\psi^j$ is a Dirac spinor and the indices
$j, k$ are summed over from $1$ to $N$, and the
gamma matrices $\gamma_\mu (\mu = 0,..., D-1)$
are defined so as to satisfy the Clifford algebra,
$\{ \gamma_\mu, \gamma_\nu \} = 2 g_{\mu\nu} {\bf 1}
= 2 {\rm diag}(1,-1,...,-1)$.
\par
By introducing an auxiliary vector field
$A_\mu$, this Lagrangian is equivalently rewritten as
\begin{eqnarray}
 {\cal L}_{Th'}
 &=& \bar \psi^j i \gamma^\mu
 (\partial_\mu - i {g \over \sqrt{N}} A_\mu )
 \psi^j - m_j \bar \psi^j \psi^j
 + {M^2 \over 2} A_\mu^2,
 \label{th'}
\end{eqnarray}
where we have introduced a parameter $M(=1)$ with the
dimension of mass, $dim[m]=dim[M]=1$ and put
$G={g^2 \over M^2}$.
Thanks to the parameter $M$, all the fields have the
corresponding canonical dimensions:
$dim[\bar \psi]=dim[\psi]=(D-1)/2,
dim[A_\mu]=(D-2)/2$ and then the coupling constant has
the dimension:
$dim[g]=(4-D)/2, dim[G]=2-D$.
\par
The theory with this Lagrangian is identical with that of
the massive vector field with which the fermion
couples minimally, since a kinetic term for $A_\mu$ is
generated through the radiative correction although it
is absent originally.
As is known from the study of massive vector boson theory
\cite{FIK90},
the  Thirring model with the Lagrangian~(\ref{th'}) is
cast into the form which is invariant under the
Becchi-Rouet-Stora (BRS) transformation by introducing
an additional field $\theta$.  The field $\theta$ is
called the St\"uckelberg field and identified with the
Batalin-Fradkin (BF) field \cite{BF87} in the general
formalism for the constrained system
\cite{BF86}. Then we start from the Lagrangian with
covariant gauge-fixing \cite{IKSY95,Kondo95a}:
\begin{eqnarray}
 {\cal L}_{Th''}
 &=& \bar \psi^j i \gamma^\mu
 (\partial_\mu - i{g \over \sqrt{N}}  A_\mu )
 \psi^j - m_j \bar \psi^j \psi^j
 + {M^2 \over 2}(A_\mu - \sqrt{N} M^{-1}
 \partial_\mu \theta )^2
 \nonumber\\&&
 - A_\mu \partial^\mu B
+ {\xi \over 2} B^2
 + i \partial^\mu \bar C \partial^\mu C.
 \label{th''}
\end{eqnarray}
Actually this Lagrangian is invariant under the BRS
transformation:
\begin{eqnarray}
 {\bf \delta}_B A_\mu(x) &=&  \partial_\mu C(x),
\nonumber\\
 {\bf \delta}_B B(x) &=& 0,
\nonumber\\
 {\bf \delta}_B C(x) &=& 0,
\nonumber\\
 {\bf \delta}_B \bar C(x) &=&  i B(x),
\nonumber\\
 {\bf \delta}_B \theta(x) &=& {M \over \sqrt{N}} C(x) ,
\nonumber\\
 {\bf \delta}_B \psi^j(x) &=&  {ig \over \sqrt{N}}
C(x) \psi^j(x),
\end{eqnarray}
where $C(x)$ and $\bar C(x)$ are ghost fields, and $B(x)$
is the Nakanishi-Lautrap Lagrange multiplier field.
\par
First we consider the case of $D \ge 3$.  The
two-dimensional case is discussed separately in section 4.
By introducing an auxiliary vector field
$f_\mu$, the "mass term" of the gauge field is linearized:
For $K_\mu = \sqrt{N} M^{-1} \partial_\mu \theta$,
\begin{eqnarray}
&& \int {\cal D}\theta \exp \left\{ i \int d^Dx
 {1 \over 2}M^2(A_\mu - K_\mu)^2 \right\}
\nonumber\\
&=& \int {\cal D}\theta \int {\cal D} f_\mu
\exp \left\{ i \int d^Dx \left[  - {1 \over 2} f_\mu
f^\mu{} + M f^\mu{} (A_\mu - K_\mu) \right] \right\}
\nonumber\\
&=&  \int {\cal D} f_\mu \delta(\partial^\mu f_\mu)
\exp \left\{ i \int d^Dx \left[  - {1 \over 2} f_\mu
f^\mu{} + M f^\mu{} A_\mu \right] \right\},
\end{eqnarray}
where in the last step we have integrated out
the scalar mode $\theta$.
\par
Applying the Hodge decomposition \cite{Wallace70}
\footnote{
Let $\omega$ be a $p$-form.  Then there is a $(p+1)$-form
$\alpha$, a $(p-1)$-form $\beta$ and a harmonic $p$-form $h$
(i.e., obeying $\delta h = 0 = dh$) such that
$$
 \omega = \delta \alpha + d \beta + h.
$$
We can restrict ourselves to the topologically trivial
space $\Omega$ for which there are no harmonic forms.
This is equivalent to saying that each $p$-form
$\omega$ obeying $d\omega=0$ is of the form $\omega =
d\beta$ (Poincare's lemma) and we say $\Omega$ has trivial
(co)homology.
{}From now on we assume that the harmonic form is absent:
$h=0$.}
to the 1-form $f_{\mu}$, we see
$f_{\mu}$  is written as
\begin{eqnarray}
f_{\mu_1} = \partial_{\mu_1} \phi
+ \epsilon_{\mu_1 ...\mu_{D}}
\partial^{\mu_2} H^{\mu_3 ...\mu_{D}},
\end{eqnarray}
where we have introduced the antisymmetric tensor field
$H_{\mu_1...\mu_{D-2}}$ of rank $D-2$, which satisfies
the Bianchi identity.
Since $f_\mu$ is divergence free, $\partial^\mu f_\mu=0$,
we can put
$
f_{\mu_1}  =   \epsilon_{\mu_1 ...\mu_{D}}
\partial^{\mu_2} H^{\mu_3 ...\mu_{D}}.
$
Then we have
\begin{eqnarray}
&& \int {\cal D}\theta  \exp \left\{ i \int d^Dx
 {1 \over 2}M^2(A_\mu - K_\mu)^2 \right\}
\nonumber\\
&=&
\int {\cal D} H_{\mu_1 ...\mu_{D-2}}
\exp \Big\{ i \int d^Dx \Big[
\nonumber\\&&
{(-1)^D \over 2(D-1)}
 \tilde H{}_{\mu_1 ...\mu_{D-1}}
\tilde H{}^{\mu_1 ...\mu_{D-1}}
+ M \epsilon^{\mu_1 \mu_2 ...\mu_{D}}
A_{\mu_1} \partial_{\mu_2} H{}_{\mu_3 ...\mu_{D}}
 \Big] \Big\},
 \label{dualrelation}
\end{eqnarray}
where we have defined
\begin{eqnarray}
\tilde H^{\mu_1 ...\mu_{D-1}}
= \partial^{\mu_1} H^{\mu_2 ...\mu_{D-1}}
- \partial^{\mu_2} H^{\mu_1\mu_3 ...\mu_{D-1}}
+ ...
+ (-1)^D \partial_{\mu_{D-1}} H^{\mu_1 ...\mu_{D-2}}.
\end{eqnarray}
This result is a generalization of \cite{DM92} for $D=3$
and coincides with the result of \cite{KL94,IKSY95}.
\par
Then we obtain the equivalent Lagrangian with the mixed
term between $A_\mu$ and $H_{\mu_1 ...\mu_{D-2}}$:
\begin{eqnarray}
 {\cal L}_{Th'''}
 &=& \bar \psi^j i \gamma^\mu D_\mu[A]
 \psi^j - m_j \bar \psi^j \psi^j
\nonumber\\&&
+ {(-1)^D \over 2(D-1)}
 \tilde H_{\mu_1 ...\mu_{D-1}}
\tilde H^{\mu_1 ...\mu_{D-1}}
+ M \epsilon^{\mu_1 ...\mu_{D}}  A_{\mu_1}
\partial_{\mu_2} H_{\mu_3 ...\mu_{D}}
\nonumber\\&&
 - A_\mu \partial^\mu B + {\xi \over 2} B^2 ,
 \label{th'''}
\end{eqnarray}
where $D_\mu[A]$ is the covariant derivative:
\begin{eqnarray}
D_\mu[A] = \partial_\mu - i {g \over \sqrt{N}} A_\mu .
\end{eqnarray}

\par
Integrating out the fermion field $\bar \psi, \psi$, we
thus obtain the bosonised action of the Thirring model:
\begin{eqnarray}
  S_B &=&
\sum_{j=1}^{N} \ln \det [i \gamma^\mu D_\mu[A] + m_j]
\nonumber\\&&
+ \int d^Dx \Biggr[ {(-1)^D \over 2(D-1)}
 \tilde H_{\mu_1 ...\mu_{D-1}}
\tilde H^{\mu_1 ...\mu_{D-1}}
+ M \epsilon^{\mu_1 ...\mu_{D}}
  A_{\mu_1}
\partial_{\mu_2} H_{\mu_3 ...\mu_{D}}
\nonumber\\&&
 - A_\mu \partial^\mu B + {\xi \over 2} B^2
 \Biggr] .
\end{eqnarray}
\par
To see the origin of the field $H_{\mu_1 ...\mu_{D-2}}$, we
integrate out the gauge field.  Then, we obtain the
partition function:
\begin{eqnarray}
Z &=&   \int {\cal D} B
\int {\cal D} \bar \psi \int {\cal D} \psi
\int {\cal D} H_{\mu_1 ...\mu_{D-2}}
\nonumber\\&&
\times \delta({g \over \sqrt{N}}\bar \psi^j  \gamma^\mu
\psi^j  - M \epsilon^{\mu \mu_2 ...\mu_{D}}
\partial_{\mu_2} H_{\mu_3 ...\mu_{D}}
-\partial^\mu B)
\nonumber\\&&
\times \exp \Big\{ i \int d^Dx \Big[
\bar \psi^j i \gamma^\mu
(\partial_\mu)
 \psi^j - m_j \bar \psi^j \psi^j
\nonumber\\&&
+ {(-1)^D \over 2(D-1)}
 \tilde H_{\mu_1 ...\mu_{D-1}}
\tilde H^{\mu_1 ...\mu_{D-1}}
 + {\xi \over 2} B^2
 \Big] \Big\}.
\end{eqnarray}
This implies that the dual field $H_{\mu_1 ...\mu_{D-2}}$
is a composite of the fermion and antifermion.
\par
The correspondence between the original Thirring model and
the bosonized theory is generalized to the correlation
function. We introduce the source $b_\mu$ for the current
\begin{eqnarray}
{\cal J}_\mu = \bar \psi^j \gamma_\mu \psi^j.
\end{eqnarray}
Adding the source term ${\cal J}_\mu b^\mu$ to the
original Lagrangian eq.~(\ref{th}):
\begin{eqnarray}
{\cal L}_{Th}[b_\mu] \equiv
{\cal L}_{Th}+{\cal J}_\mu b^\mu,
\end{eqnarray}
we obtain
\begin{eqnarray}
 {\cal L}_{Th'}[b_\mu]
 &=& \bar \psi^j i \gamma^\mu
 (\partial_\mu - i {g \over \sqrt{N}} A_\mu -i b_\mu)
 \psi^j - m_j \bar \psi^j \psi^j
 + {M^2 \over 2}(A_\mu)^2 .
 \label{th'b}
\end{eqnarray}
After introducing the BF field $\theta$ and shifting
$A_\mu \rightarrow A_\mu - {\sqrt{N} \over g} b_\mu$,
we obtain
\begin{eqnarray}
 {\cal L}_{Th''}[b_\mu]
 &=& \bar \psi^j i \gamma^\mu
 (\partial_\mu - i{g \over \sqrt{N}}  A_\mu )
 \psi^j - m_j \bar \psi^j \psi^j
+ {M^2 \over 2}(A_\mu - {\sqrt{N} \over g} b_\mu - K_\mu )^2
 \nonumber\\&&
 - A_\mu \partial^\mu B
+ {\xi \over 2} B^2
 + i \partial^\mu \bar C \partial_\mu C .
 \label{th''b}
\end{eqnarray}
Repeating the same steps as before, we arrive at
\begin{eqnarray}
 {\cal L}_{Th'''}[b_\mu]
 = {\cal L}_{Th'''}
+ {\sqrt{N} \over g}
M \epsilon^{\mu_1 ...\mu_{D}}  b_{\mu_1}
\partial_{\mu_2} H_{\mu_3 ...\mu_{D}} .
 \label{th'''b}
\end{eqnarray}
This leads to the equivalence of the partition function
in the presence of the source $b_\mu$:
\begin{eqnarray}
Z_{Th}[b_\mu]=Z_{Bosonised}[b_\mu].
\end{eqnarray}
Therefore the
connected correlation function has the following
correspondence between the Thirring model and the
bosonised theory with the action $S_B$:
\begin{eqnarray}
 \langle {\cal J}_{\mu_1}; ... ;{\cal J}_{\mu_n}
\rangle_{Th} = \langle \eta_{\mu_1}; ... ; \eta_{\mu_n}
 \rangle_{Bosonised},
 \label{correspond}
\end{eqnarray}
where
\begin{eqnarray}
\eta^{\mu_1}
= {1 \over \sqrt{G/N}}  \epsilon^{\mu_1 ...\mu_{D}}
\partial_{\mu_2} H_{\mu_3...\mu_{D}} .
\end{eqnarray}
In the following we discuss how to integrate out the
auxiliary field $A_\mu$ to obtain the bosonic theory which
is written in terms of the field $H_{\mu_1 ...\mu_{D-2}}$
only.

\section{(2+1)-dimensional case}
\setcounter{equation}{0}
In the three-dimensional case,
$D=3$,
\begin{eqnarray}
 S_B &=&
\sum_{j=1}^{N} \ln \det [i \gamma^\mu D_\mu[A] + m_j]
\nonumber\\&&
+ \int d^3 x \Big[ - {1 \over 4}
 \tilde H_{\mu \nu} \tilde H^{\mu \nu}
+ M \epsilon^{\mu \nu \rho}
  A_{\mu}  \partial_{\nu} H_{\rho}
 - A_\mu  \partial^\mu B  + {\xi \over 2} B^2 \Big],
 \label{effaction}
\end{eqnarray}
where
\begin{eqnarray}
\tilde H_{\mu\nu} = \partial_\mu H_\nu - \partial_\nu H_\mu.
\end{eqnarray}
The Matthews-Salam determinant in eq.~(\ref{effaction}) is
calculated with the aid of appropriate regularizations.
There are various gauge-invariant regularization methods:
1) Pauli-Villars \cite{DJT81,Redlich84,KS85,Alvarez90},
2) lattice \cite{So85,CL89},
3) analytic \cite{PST92},
4) dimensional \cite{Martin90,DW93},
5) Zavialov \cite{Novotny92},
6) parity-invariant Pauli-Villars (variant of chiral gauge
invariant Pauli-Villars by Frolov and Slavnov)
\cite{Kimura94},
7) high covariant derivative \cite{AF90},
8) zeta-function \cite{Gamboa85}.
However it should be remarked that the methods 1) and 2)
give regulator dependent result for the Chern-Simons part.
Appropriate choice of regularization leads to the
regulator independent result, see for example
\cite{Kimura94}.
In the massless limit, $m \rightarrow 0$, it is shown
\cite{Redlich84} up to one-loop that
\begin{eqnarray}
&&  \ln \det [i \gamma^\mu  (\partial_\mu - i{g \over
\sqrt{N}} A_\mu) + m]
\nonumber\\
&\rightarrow& {sgn(m) \over N} {i \over 16\pi}g^2 \int d^3x
\epsilon^{\mu\nu\rho} F_{\mu\nu} A_\rho +  I_{PC}[A_\mu],
\end{eqnarray}
where $sgn(m)$ denotes the signature of $m$ and the
parity-conserving term is given by
\begin{eqnarray}
 I_{PC}[A_\mu] = {1 \over \pi^2} \zeta({3 \over 2})
 \int d^3 x
 \left( {g \over 2\sqrt{N}} F_{\mu\nu}^2 \right)^{3/2}.
\end{eqnarray}
Therefore the bosonization of the massless Thirring model
in (2+1) dimensions would lead to highly complicated
bosonic theory.
\par
In what follows we consider the  large fermion mass limit,
$m \rightarrow \infty$.
For the $2 \times 2$ gamma matrices corresponding to the
two-component fermion, $\bar \psi, \psi$,
\begin{eqnarray}
&&  N \ln \det [i \gamma^\mu D_\mu[A] + m_j]
- N \ln \det [i \gamma^\mu \partial_\mu  + m_j]
\nonumber\\
&=& N \Tr \ln \left[ 1 + (i \gamma^\mu \partial_\mu  +
m_j)^{-1} {g \over \sqrt{N}} \gamma^\mu A_\mu \right]
\nonumber\\
&=& sgn(m_j) {ig^2 \over 16\pi} \int d^3x
\epsilon^{\mu\nu\rho} F_{\mu\nu} A_\rho
 - {g^2 \over 24\pi |m_j|} \int d^3x F_{\mu\nu}F^{\mu\nu}
+ {\cal O}\left({\partial^2 \over |m_j|^2} \right),
\label{det}
\end{eqnarray}
where
$sgn(m)$ denotes signature of the fermion mass $m$,
$sgn(m)=m/|m|$. This is understood as follows.  Note that
\begin{eqnarray}
 N \Tr \ln \left[ 1 + (i \gamma^\mu \partial_\mu  +
m)^{-1} {g \over \sqrt{N}} \gamma^\mu A_\mu \right]
=  {1 \over 2} \int d^Dx A^\mu(x)
\Pi_{\mu\nu}(\partial;m) A^\nu(x)
+ ....
\end{eqnarray}
For $D=3$, the vacuum polarization tensor is given by
\begin{eqnarray}
\Pi_{\mu\nu}(\partial;m) = \left( g_{\mu\nu} -
 {\partial_\mu \partial_\nu \over \partial^2}  \right)
\Pi_T(-\partial^2,m)
+ i\epsilon_{\mu\nu\rho} \partial^\rho
\Pi_O(-\partial^2;m),
\end{eqnarray}
where
\begin{eqnarray}
 \Pi_T(k^2;m)
 = -  {g^2 \over 2\pi} k^2
 \int_0^1 d\alpha {\alpha(1-\alpha) \over
 [m^2-\alpha(1-\alpha)k^2]^{1/2}},
\end{eqnarray}
and
\begin{eqnarray}
 \Pi_O(k^2;m)
 = -  {g^2m \over 4\pi}
 \int_0^1 d\alpha {1 \over
[m^2-\alpha(1-\alpha)k^2]^{1/2}}.
\end{eqnarray}
In the large fermion mass limit, we obtain eq.~(\ref{det}).

\par
Thus the Thirring model in the large fermion mass limit
is equivalent to the bosonized theory
with the interpolating Lagrangian:
\begin{eqnarray}
 {\cal L}_I[A_\mu,H_\mu] &=&   - {1 \over 4}
 \tilde H_{\mu \nu} \tilde H^{\mu \nu}
+ {M \over 2} \epsilon^{\mu \nu \rho}
   \tilde H_{\mu \nu} A_{\rho}
  + {i \theta_{CS} \over 4}
\epsilon^{\mu\nu\rho} F_{\mu\nu} A_\rho
\nonumber\\&&
 - {1 \over N} \sum_{j=1}^{N}
 {GM^2 \over 24\pi |m_j|} F_{\mu\nu}F^{\mu\nu}
 - {1 \over 2\xi} (\partial^\mu A_\mu)^2
 + {\cal O}\left({\partial^2 \over m_j^2} \right),
 \label{master}
\end{eqnarray}
where
\begin{eqnarray}
 \theta_{CS} = {1 \over N} \sum_{j=1}^{N} sgn(m_j)
 {GM^2 \over 4\pi}.
\end{eqnarray}
Integrating out the $\tilde H_{\mu\nu}$ field via
$
H_\rho^* := {1 \over 2}\epsilon_{\mu \nu \rho} \tilde
H^{\mu\nu} = \epsilon_{\mu\nu\rho} \partial^\mu H^\nu
$
in the interpolating Lagrangian, we obtain the
self-dual Lagrangian in the same sense as used in Deser and
Jackiw
\cite{DJ84}:
\begin{eqnarray}
 {\cal L}_{SD}[A_\mu] &=&   {M^2 \over 2} A^\mu A_\mu
  + {i \theta_{CS} \over 4}
\epsilon^{\mu\nu\rho} F_{\mu\nu} A_\rho,
 \label{sd}
\end{eqnarray}
up to the lowest order of $1/m$.
This implies that the Thirring model is equivalent to the
self-dual model with the Lagrangian ${\cal L}_{SD}$ in
the lowest order of $1/m$.
\par
We notice that the interpolating Lagrangian we have just
obtained is essentially equivalent to the master Lagrangian
of Deser and Jackiw \cite{DJ84}.
The master Lagrangian for $A_\mu$ and $H_\mu{}^{*}$  is
given by
\begin{eqnarray}
 {\cal L}_{DJ} =  {1 \over 2} A^\mu A_\mu
 - A_\mu H^\mu{}^{*}
 + {1 \over 2} \tilde m H^\mu{}^{*} H_\mu .
\end{eqnarray}
Note that the role of the auxiliary field $A_\mu$ and the
new field $H_\mu$ is interchanged in our interpolating
Lagrangian compared with that in \cite{FS94} based on the
master Lagrangian of Deser and Jackiw.
Hence the integration over the auxiliary field is
non-trivial in our interpolating Lagrangian.
\par
If $m_j=m$ for all $j=1,...,N$, then the original
Lagrangian~(\ref{th}) has $O(N)$ symmetry.  If we adopt the
fermion mass term such that
\begin{equation}
  m_j = \cases{ m & ($j=1,...,N-k$) \cr
               -m & ($j=N-k+1,...,N$)},
 \label{pattern}
\end{equation}
the Lagrangian has $O(N-k) \times O(k)$ symmetry.
For this mass term, we obtain
\begin{eqnarray}
\theta_{CS}
= \left(1-2{k \over N} \right) {GM^2 \over 4\pi}.
\label{cscoeff}
\end{eqnarray}
Hence, if we take $k=N/2$ in the fermion mass term,
the theory has $O(N/2) \times O(N/2)$ symmetry and
$\theta_{CS}=0$ follows, although the kinetic term for the
field $A_\mu$ in the next-to-leading order is unchanged.
See the Vafa-Witten argument \cite{VW84b}.
A similar situation occurs in the formulation
which uses the 4-component fermion  with $4 \times 4$
gamma matrices where the   Chern-Simons term in the
determinant disappear, since
$\tr(\gamma_\mu \gamma_\nu \gamma_\rho) = 0$.
In this case, the Thirring model is equivalent to the
bosonised theory with the Lagrangian:
\begin{eqnarray}
 {\cal L}_{I'} &=&   - {1 \over 4}
 \tilde H_{\mu \nu} \tilde H^{\mu \nu}
+ M \epsilon^{\mu \nu \rho}
  A_{\mu}  \partial_{\nu} H_{\rho}
 - {g^2 \over 24\pi |m|} F_{\mu\nu}F^{\mu\nu}
\nonumber\\&&
 - {1 \over 2\xi} (\partial^\mu A_\mu)^2
  + {\cal O}\left({\partial^2 \over |m|^2} \right).
\end{eqnarray}
Such a case is discussed in the final section.
\par
Now we proceed to the step of integrating out the gauge
field  $A_\mu$, assuming $\theta_{CS}\not=0$ and the
fermion mass pattern~(\ref{pattern}), i.e., $|m_j|=m$ for
all $j$. By defining
\begin{eqnarray}
 J^\mu(x) =
 M \epsilon^{\mu\nu\rho} \partial_\nu H_\rho(x)
 = {M \over 2} \epsilon^{\mu \nu \rho}
   \tilde H_{\mu \nu}(x) ,
\end{eqnarray}
and
\begin{eqnarray}
K^{\mu \nu}(x,y)
 = \left[
 i \theta_{CS} \epsilon^{\mu\nu\rho}\partial_\rho
 + {1 \over \xi} \partial^\mu \partial^\nu
 + {g^2 \over 6\pi |m|} (\partial^2 g^{\mu\nu}
 -\partial^\mu \partial^\nu) \right] \delta(x-y),
\end{eqnarray}
the interpolating action is written in the form:
\begin{eqnarray}
 S_I &=&   \int d^3x \left[ - {1 \over 4}
 \tilde H_{\mu \nu} \tilde H^{\mu \nu}
 + A_\mu(x) J^\mu(x) \right]
\nonumber\\&&
 +  \int d^3x \int d^3y
 {1 \over 2} A_\mu(x) K^{\mu\nu}(x,y) A_\nu(y)
 + {\cal O}\left({\partial^2 \over m^2} \right).
 \label{masteraction}
\end{eqnarray}
The  $A_\mu$ integration in the interpolating action  is
performed  by using
\begin{eqnarray}
&& \int {\cal D} A_\mu \exp \left\{ i
\int d^3x \int d^3y
 {1 \over 2} A_\mu(x) K^{\mu\nu}(x,y) A_\nu(y)
 + i \int d^3x A_\mu(x) J^\mu(x)
 \right\}
\nonumber\\ &=&  \exp \left\{ i \int d^3x \int
d^3y
 {1 \over 2} J^\mu(x) K^{-1}_{\mu\nu}(x,y) J^\nu(y)
 \right\},
\end{eqnarray}
up to a constant which is independent of the
field variable.
Here the inverse is obtained as
\begin{eqnarray}
 K^{-1}_{\mu \nu}(x,y)
 &=& {i \over \theta_{CS}}
 \epsilon_{\mu\nu\rho} \partial^\rho  \Delta^{(1)}(x,y)
 +  \xi  \partial_\mu \partial_\nu \Delta^{(2)}(x,y)
\nonumber\\&&
 + {g^2 \over 6\pi \theta_{CS}^2 |m|} \left( g_{\mu\nu}
 - \partial_\mu \partial_\nu \Delta^{(1)}(x,y)   \right)
 + {\cal O}\left({1 \over |m|^2} \right),
\end{eqnarray}
where $\Delta^{(1)}=1/\partial^2$ and
$\Delta^{(2)}=1/\partial^4$ in the sense
$\partial^2 \Delta^{(1)}(x,y)=\delta(x-y)$ and
$(\partial^2)^2 \Delta^{(2)}(x,y)=\delta(x-y)$.
By using this formula,
$A_\mu$ integration is performed:
\begin{eqnarray}
J^\mu K^{-1}_{\mu\nu} J^\nu
= {M^2 \over i \theta_{CS}} \epsilon^{\mu \nu \rho}
  H_{\mu}  \partial_{\nu} H_{\rho}
  + {g^2 M^2 \over 12\pi \theta_{CS}^2 |m|}
  \tilde H_{\mu\nu}^2
   + {\cal O}\left({1 \over |m|^2} \right),
\end{eqnarray}
which is independent of the gauge-fixing parameter $\xi$
in the original theory.
\par
Thus we arrive at an effective bosonised Lagrangian for the dual
field $H_\mu$ alone:
\begin{eqnarray}
 {\cal L}_{MCS} &=& - {1 \over 4} \left( 1 +
{g^2 M^2 \over 6\pi \theta_{CS}^2 |m|} \right)
 \tilde H_{\mu \nu} \tilde H^{\mu \nu}
+   {iM^2 \over 2\theta_{CS}} \epsilon^{\mu\nu\rho}
  H_{\mu}  \partial_{\nu} H_{\rho}
\nonumber\\
&& + {\cal O}\left({\partial^2 \over |m|^2} \right).
\end{eqnarray}
Note that the gauge-parameter dependence has dropped out
in the bosonized theory.
In the interpolating Lagrangian ${\cal L}_I$ the gauge
degree of freedom for the $A_\mu$ field is fixed by the
gauge-fixing term
${1 \over 2\xi}(\partial^\mu A_\mu)^2$.
However there is an additional gauge symmetry for the new
field $H_\mu$: the Lagrangian ${\cal L}_I$ is invariant
under the gauge transformation
$\delta H_\mu=\partial_\mu \omega$ independently of
$A_\mu$, which leads to $\delta f_\mu=0$ in the master
Lagrangian. Therefore we must add a gauge-fixing term for
the $H_\mu$ field to the bosonized Lagrangian
${\cal L}_{MCS}$.

\par
Thus, to the leading order of $1/m$ expansion, the
Thirring model partition function coincides with that of
the Maxwell-Chern-Simons theory. This result agrees with
that in \cite{FS94} obtained up to the leading of $1/m$
where the less direct procedure is adopted to show this
equivalence using the self-dual action by way of the
interpolating action.  Up to the next-to-leading order of
$1/m$, we have shown that the equivalence between the low
energy sector of a theory of 3-dimensional fermions
interacting via a current-current term and gauge bosons
with Maxwell-Chern-Simons term is preserved.   The Thirring
spin-1/2 fermion with the Thirring coupling $g^2/N$ is
equal to a spin-1 massive excitation with mass
\begin{eqnarray}
{\pi \over g^2}
\left( 1 - {g^2 M^2 \over 6\pi \theta_{CS}^2 |m|}
\right) + {\cal O}({1 \over m^2}),
\end{eqnarray}
in 2+1 dimensions.
In 2+1 dimensions there is the following correspondence
between the  original Thirring model and the bosonized
theory:
\begin{eqnarray}
\bar \psi^j \gamma^\mu \psi^j \leftrightarrow
 {1 \over \sqrt{G/N}} \epsilon^{\mu\nu\rho}
\partial_{\nu} H_{\rho} .
\end{eqnarray}

\par
Especially, for $D=3$, the relation Eq~(\ref{dualrelation})
shows that the London action (without a kinetic term
$F_{\mu\nu}^2$) for superconductivity:
\begin{eqnarray}
 {\cal L}
 =  {1 \over 2}(\partial_\mu \phi - M A_\mu)^2
 =  {M^2 \over 2}(A_\mu - M^{-1} \partial_\mu \phi)^2,
\end{eqnarray}
is equivalent to the Chern-Simons term:
\begin{eqnarray}
 {\cal L} = - {1 \over 4} H_{\mu\nu} H^{\mu\nu}
 - M \epsilon^{\mu\nu\rho} A_\mu \partial_\nu H_\rho.
\end{eqnarray}
This fact was already pointed out in \cite{DM92}.
The missing kinetic term for $A_\mu$ is generated through
the radiative correction as shown above.
The Meissner effect in superconductivity is nothing
but the Higgs phenomenon: the photon (massless gauge
field) becomes massive gauge boson by absorbing the massless
Nambu-Goldstone boson (scalar mode). The mixed Chern-Simons
action does not break the parity in sharp contrast to the
ordinary Chern-Simons term.  Therefore this model may be a
candidate for the high-$T_c$ superconductivity without
parity violation, as suggested in
\cite{DM92}.
\par
Finally we wish to point out that, in the large $m$ limit,
the Thirring model is also equivalent to the
Chern-Simons-Higgs model up to the leading order of $1/m$
and to the Maxwell-Chern-Simons-Higgs model up to the
next-to-leading order of $1/m$, since
\begin{eqnarray}
 {\cal L}_{CSH}
 &=&  (D_\mu \varphi)^\dagger (D^\mu \varphi)
 +  {i\theta_{CS} \over 4} \int d^3x
\epsilon^{\mu\nu\rho} F_{\mu\nu} A_\rho
 \nonumber\\&&
 - {g^2 \over 24\pi |m|} \int d^3x F_{\mu\nu}F^{\mu\nu} +
{\cal O}\left({\partial^2 \over |m|^2} \right),
\end{eqnarray}
apart from the gauge-fixing term.
This result is consistent with the assertion of
Deser and Yang \cite{DY89}.

\section{(1+1)-dimensional case}
\setcounter{equation}{0}
The bosonization of the (1+1)-dimensional Thirring model
has a long history and is well known.  Therefore there are
a lot of references on bosonization which can not be
exhausted, see for example
\cite{Klaiber68,Coleman75,Halpern75,Witten84}
based on the canonical formalism and
\cite{GSS81,PW83,VDP84,Dorn86,Rothe86,DV88}
based on the path integral formalism.
In this section we reproduce the well-known result based
on our method.
\par
 For $D=2$, it is easy to show that the
bosonized action is given by
\begin{eqnarray}
  S_B &=&
N \ln \det [i \gamma^\mu D_\mu[A] + m]
\nonumber\\&&
+ \int d^2x \Biggr[ {1 \over 2}
  (\partial_\mu H )^2
+ M \epsilon^{\mu\nu}  A_{\mu}  \partial_{\nu} H
 - A_\mu \partial^\mu B + {\xi \over 2} B^2
 \Biggr] .
\end{eqnarray}

The determinant is calculated as
\begin{eqnarray}
&&
N \ln {\det [i \gamma^\mu D_\mu[A] + m] \over
 \det [i \gamma^\mu \partial_\mu  + m]}
\nonumber\\
&=&   {1 \over 2} \int d^2x
  A^\mu(x) \left( g_{\mu\nu} - {\partial_\mu  \partial_\nu
\over \partial^2}  \right) \Pi_T(-\partial^2;m) A^\nu(x) ,
\end{eqnarray}
where
\begin{eqnarray}
 \Pi_T(k^2;m) = - {g^2 \over \pi} k^2 \int_0^1 d \alpha
 {\alpha(1-\alpha) \over m^2-\alpha(1-\alpha)k^2}.
\end{eqnarray}
\par
In the massless case, $m=0$,
\begin{eqnarray}
\Pi_T(k^2;m=0) =  {g^2 \over \pi},
\end{eqnarray}
and hence
\begin{eqnarray}
K_{\mu \nu}
 =  {g^2 \over \pi} \left( g_{\mu\nu} -
 {\partial_\mu  \partial_\nu \over \partial^2}  \right)
 + {1 \over \xi} \partial_\mu \partial_\nu ,
\end{eqnarray}
whose inverse is given by
\begin{eqnarray}
 K^{-1}_{\mu \nu}
 = {\pi \over g^2}
 \left( g_{\mu\nu}
 - {\partial_\mu  \partial_\nu \over \partial^2}  \right)
+ \xi  {\partial_\mu  \partial_\nu \over \partial^4} .
\end{eqnarray}
Putting
\begin{eqnarray}
 J^\mu = M \epsilon^{\mu\nu} \partial_\nu H,
\end{eqnarray}
and integrating out the $A_\mu$ field, we obtain the
bosonised theory:
\begin{eqnarray}
  S_B &=&
 \int d^2x   {1 \over 2}
 \left( 1 + {\pi \over G} \right)
  (\partial_\mu H )^2  .
\end{eqnarray}
The massless Thirring model in two dimensions is equivalent
to  the massless scalar field theory, as long as the
action is bounded from below, i.e.,
$G:=g^2/M^2>0$ or $G<-\pi $.
For more details on the massless case, see reference
\cite{IKSY95}.
\par
On the other hand, in the large $m$ limit, we find
\begin{eqnarray}
\Pi_T(k^2;m) = - {g^2 \over \pi}
\left[{1 \over 6}{k^2 \over m^2}
+ {1 \over 30} {k^4 \over m^4} \right]
+ {\cal O}({k^4 \over m^4}),
\end{eqnarray}
and hence
\begin{eqnarray}
K_{\mu \nu}
 =  {g^2 \over 6\pi} {\partial^2 \over m^2}
 \left( 1 - {1 \over 5}{\partial^2 \over m^2} \right)
 \left( g_{\mu\nu} -
 {\partial_\mu  \partial_\nu \over \partial^2}  \right)
 + {1 \over \xi} \partial_\mu \partial_\nu
 + {\cal O}({\partial^2 \over m^2}) .
\end{eqnarray}
The inverse is obtained as
\begin{eqnarray}
 K^{-1}_{\mu \nu}
 = {6\pi \over g^2} {m^2 \over \partial^2}
 \left( 1 + {1 \over 5}{\partial^2 \over m^2} \right)
 \left( g_{\mu\nu}
 - {\partial_\mu  \partial_\nu \over \partial^2}  \right)
+ \xi  {\partial_\mu  \partial_\nu \over \partial^4}
+ {\cal O}({\partial^2 \over m^2}) .
\end{eqnarray}
Hence we obtain the bosonized theory after
integrating out the $A_\mu$ field:
\begin{eqnarray}
 S_B &=&
 \int d^2x   \left[
 {1 \over 2}
 \left( 1 + {6 \over 5}{\pi \over G}  \right)
 (\partial_\mu H )^2
  - {3\pi m^2 \over G}  H^2 \right]
  + {\cal O}({\partial^4 \over m^2})  .
\end{eqnarray}
Thus in two dimensions the massive Thirring model with
large mass $m \gg 1$ is equivalent to the scalar field
theory with large mass:
\begin{eqnarray}
 m_H = \sqrt{{6\pi m^2 \over G}
\left( 1 + {6 \over 5}{\pi \over G} \right)}.
\end{eqnarray}
In the massive limit, the Thirring model is physically
sensible for $G>0$.
\par
In two dimensions there is a correspondence between the
Thirring model and the bosonised theory:
\begin{eqnarray}
\bar \psi^j \gamma^\mu \psi^j  \leftrightarrow
{1 \over \sqrt{G/N}} \epsilon^{\mu\nu}   \partial_{\nu} H .
\label{corr}
\end{eqnarray}
\par
The above results are reasonable as shown in the
following.
The massive Thirring model is not exactly soluble even
in (1+1)-dimensions.
However the (1+1)-dimensional massive Thirring model is
equivalent to the sine-Gordon model
\cite{Coleman75}:
\begin{eqnarray}
 {\cal L}_{sG}
 = {1 \over 2} \partial_\mu \varphi \partial^\mu \varphi
 + {\alpha \over \beta^2} [\cos (\beta \varphi) -1 ],
\end{eqnarray}
if the following identifications are made between the two
theories:
\begin{eqnarray}
 1 + {G \over \pi} &=& {4\pi \over \beta^2},
 \label{c}
 \\
 \bar \psi \gamma^\mu \psi
 &=& - {\beta \over 2\pi} \epsilon^{\mu\nu}
\partial_\nu \varphi,
\label{current}
\\
m \bar \psi \psi &=& - {\alpha \over \beta^2}
\cos (\beta \varphi),
\label{operator}
\end{eqnarray}
where a constant in the Lagrangian
${\cal L}_{sG}$ is adjusted so that the minimum of
the energy density is zero.
\par
The massless limit $m  \rightarrow 0$ of the massive
Thirring model corresponds to the limit
$\alpha \rightarrow  0$ in the
sine-Gordon model, i.e., a massless scalar field theory
in agreement with the above result.
On the other hand, the massive limit $m \rightarrow \infty$
corresponds to the limit $\beta \rightarrow 0$
(or $\alpha \rightarrow \infty$) which
inevitably leads to the limit $G \rightarrow \infty$ as
${\beta \over 2\pi} \sim {1 \over \sqrt{G}}$.  Hence in
this limit the Lagrangian reduces to
\begin{eqnarray}
 {\cal L}_{sG}
 = {1 \over 2} \partial_\mu \varphi \partial^\mu \varphi
 - {\alpha \over 2} \varphi^2
 + {\cal O}(\alpha \beta^2 \varphi^4).
\end{eqnarray}
Moreover the eq.~(\ref{current}) recovers the above
correspondence relation eq.~(\ref{corr}) for $N=1$.
Therefore the very massive limit $m \gg 1$ of
the 2-dimensional Thirring model is equivalent to the
massive free scalar field theory with mass $\sqrt{\alpha}$
which is identified with $m_H$ given above.

\section{Conclusion and Discussion}
\setcounter{equation}{0}
In this paper we have investigated the bosonization of the
multiflavor massive Thirring model in $D = d+1 \ge 2$
dimensions, starting from a reformulation of the Thirring
model as a gauge theory.
\par
In (1+1) dimensions, we have reproduced
the well known result \cite{Klaiber68,Coleman75,Halpern75}.
In (2+1) dimensions we have found a novel interpolating
Lagrangian:
\begin{eqnarray}
 {\cal L}_I[A_\mu,H_\mu] =   - {1 \over 4}
 \tilde H_{\mu \nu} \tilde H^{\mu \nu}
+ {M \over 2} \epsilon^{\mu \nu \rho}
   \tilde H_{\mu \nu} A_{\rho}
+ {\cal L}_G[A_\mu],
\label{masterl1}
\end{eqnarray}
where ${\cal L}_G[A_\mu]$ is the Lagrangian for the
gauge field $A_\mu$ generated from the fermion determinant
up to the next-to-leading order in $1/|m_j|$:
\begin{eqnarray}
 {\cal L}_G[A_\mu] =
 {i \theta_{CS} \over 4}
 \epsilon^{\mu\nu\rho} A_\rho F_{\mu\nu}
 - {g^2 \over 24\pi |m|} F_{\mu\nu} F^{\mu\nu}
 + {\cal O}\left({\partial^2 \over |m|^2} \right).
\end{eqnarray}
This interpolating Lagrangian is shown to interpolate
between the massive Thirring model with the Lagrangian
${\cal L}_{Th}$ and the Maxwell-Chern-Simons theory with the
Lagrangian:
\begin{eqnarray}
 {\cal L}_{MCS}[H_\mu] = - {1 \over 4} \left( 1 +
{G \over 6\pi \theta_{CS}^2 |m|} \right)
 \tilde H_{\mu \nu} \tilde H^{\mu \nu}
+   {i M^2 \over 2\theta_{CS}}  \epsilon^{\mu \nu \rho}
  H_{\mu}  \partial_{\nu} H_{\rho} .
\end{eqnarray}
In contrast to the previous interpolating Lagrangian
\cite{DJ84,FS94}, our interpolating Lagrangian is invariant
under the independent gauge transformations for two gauge
fields,
$A_\mu, H_\mu$:
\begin{eqnarray}
\delta A_\mu = \partial_\mu \lambda,
\quad
\delta H_\mu = \partial_\mu \omega.
\label{symm}
\end{eqnarray}
It is interesting to see that the weak coupling limit $G
\downarrow 0$ (i.e., $\theta_{CS} \downarrow 0$) of the
massive Thirring model is nothing but the topological
theory, the Chern-Simons theory, in the leading order of
$1/|m|$.
\par
In the multiflavor Thirring case we have treated in this
paper, a specific situation such that $\theta_{CS}=0$ may
occur for even number of 2-component fermions or
4-component fermions which is important in discussing
the chiral symmetry breaking \cite{Kondo95a} (see
eq.~(\ref{cscoeff})), in contrast to the case considered
in \cite{FS94}. In such a case, the inverse of $K_{\mu\nu}$
is given by
\begin{eqnarray}
 K^{-1}_{\mu \nu}(x,y)
 =  \xi  \partial_\mu \partial_\nu \Delta^{(2)}(x,y)
 + {3\pi |m| \over G} \Delta^{(1)}\left( g_{\mu\nu}
 - \partial_\mu \partial_\nu \Delta^{(1)} \right)(x,y).
\end{eqnarray}
This leads to the non-local Maxwell-like Lagrangian.
In the multiflavor case, therefore, the exact bosonization
of the massive Thirring model up to the next-to-leading
order in $1/m$ is possible only when fermion masses are
configurated such that the coefficient
$\theta_{CS}$ is nonzero.
\par
In order to formulate the Thirring model as a gauge theory
based on the BFV formalism, we need a kinetic term for the
gauge field $A_\mu$.  Therefore the existence of the
next-to-leading order term in ${\cal L}_G[A]$ is
essential in our formulation.
The Chern-Simons theory and the Maxwell theory has
completely different structure as a constraint system in
the bosonization scenario.   From this point of view, we
consider the bosonization in a subsequent
paper \cite{IKN95}.

\par
In dimensions $D \ge 4$, the Lagrangian ${\cal L}_G[A]$ for
the gauge field $A_\mu$ coming from integrating out the
fermion field becomes non-local, since
\begin{eqnarray}
 \Pi_T(k^2;m)
&=& - k^2 {2\tr(1) \Gamma(2-D/2) \over (4\pi)^{D/2}}
\int_0^1 d\alpha {\alpha(1-\alpha) \over
[\alpha(1-\alpha)k^2+m^2]^{2-D/2}}
\nonumber\\
&=& -  {\tr(1) \Gamma(2-D/2) \over 3(4\pi)^{D/2}}
{k^2 \over m^{4-D}}
{}_2F_1(2,2-{D \over 2},{5 \over 2};-{k^2 \over 4m^2}),
\end{eqnarray}
where ${}_2F_1(a,b,c;z)$ is a hypergeometric function.
This makes the bosonization in
the exact sense rather difficult together with the
reducibleness of the antisymmetric tensor gauge theory
written in terms of $H_{\mu_{3}...\mu_{D}}$. This case will
be discussed elsewhere.
\par
Finally we discuss how to extend our method into the
non-Abelian case.  First of all, we observe that the
following master Lagrangian is equivalent
to the interpolating Lagrangian~(\ref{th''}) apart from the
gauge-fixing and the ghost terms:
\begin{eqnarray}
 {\cal L}_M[A_\mu,H_\mu,K_\mu] =
 {M^2 \over 2}(A_\mu - K_\mu)^2
 + {1 \over 2} \epsilon^{\mu_{1}...\mu_{D}}
H_{\mu_{3}...\mu_{D}} F_{\mu_{1}\mu_{2}}[K]
  + {\cal L}_G[A_\mu] ,
\label{masterl2}
\end{eqnarray}
where $F_{\mu\nu}[K]$ is the field strength for $K_\mu$
defined by
$
F_{\mu\nu}[K] := \partial_\mu K_\nu - \partial_\nu K_\mu .
$
By introducing the vector field $f_\mu$ as in section 2,
this Lagrangian is cast into
\begin{eqnarray}
 {\cal L}_M =
 - {1 \over 2}(f_\mu)^2 + M f^\mu (A_\mu - K_\mu)
 + {1 \over 2} \epsilon^{\mu_{1}...\mu_{D}}
H_{\mu_{3}...\mu_{D}} F_{\mu_{1}\mu_{2}}[K]
  + {\cal L}_G[A_\mu]  .
\end{eqnarray}
Integrating out the field $K_\mu$, we obtain the constraint
$
f^{\mu_1} = \epsilon^{\mu_1...\mu_D} \partial_{\mu_2}
H_{\mu_{3}...\mu_{D}} .
$
Therefore the Lagrangian~(\ref{masterl2})
reproduces the interpolating Lagrangian~(\ref{th'''}) for
arbitrary dimension and especially (\ref{masterl1}) for
$D=3$.
\par
Next, we point out that,
by introducing the scalar field
\begin{eqnarray}
 \varphi = \sqrt{N} {V \over \sqrt{2}} e^{i \theta/V},
 \quad
 (V = M/g = 1/\sqrt{G}),
\end{eqnarray}
and integrating out the fermion field,  the Thirring model
as a gauge theory can be regarded with the gauged
non-linear sigma model:
\begin{eqnarray}
 {\cal L}_{H}
 =    (D_\mu[A] \varphi)^\dagger (D^\mu[A] \varphi)
+ {\cal L}_G[A] ,
\end{eqnarray}
with a local constraint:
$\varphi(x) \varphi^*(x) = {N \over 2G}$.
Actually, after redefinition of the field variable,
another form of the master Lagrangian~(\ref{masterl2}) is
obtained:
\begin{eqnarray}
 {\cal L}_M[A_\mu,H_\mu,V_\mu] =
 {M^2 \over 2}(V_\mu)^2
 + {1 \over 2} \epsilon^{\mu_{1}...\mu_{D}}
H_{\mu_{3}...\mu_{D}} F_{\mu_{1}\mu_{2}}[V+A]
  + {\cal L}_G[A_\mu] ,
\label{masterl3}
\end{eqnarray}
which is shown to be at least classically equivalent to the
non-linear sigma model with the Lagrangian ${\cal L}_H$.
Indeed the master Lagrangian~(\ref{masterl3}) has the local
gauge invariance~(\ref{symm}).
It is easy to extend the master Lagrangian~(\ref{masterl2})
or (\ref{masterl3}) into the non-Abelian case.  However the
quantum nature of the theory produces subtle problems in
bosonization.  The bosonization of a non-Abelian version of
the massive Thirring model will be given in a subsequent
paper \cite{IKN95}.

\vskip 1cm
\section*{Acknowledgments}

The author would like to thank
Atsushi Nakamura, Tadahiko Kimura, Kenji Ikegami and Toru
Ebihara for valuable discussions.
Thanks are also due to Yoonbai Kim and Koichi Yamawaki for
suggestions on the revision of the manuscript and Fidel
Schaposnik for some comments as well as technical remarks.
This work is supported in part by the Grant-in-Aid for
Scientific Research from the Ministry of Education, Science
and Culture (No.07640377).

\section*{Note added}
Quite recently, a non-Abelian version of the massive
Thirring model in (2+1)-dimensions was bosonized in
\cite{BFMS95} by following the same strategy as the
Abelian case \cite{FS94}.
After having submitted this paper for publication, the
author was informed
that the basic idea of introducing the gauge degrees of
freedom was already introduced by Burgess et al
\cite{BQ94}, although he was not aware of existence of such
works. After our manuscript has been submitted to the
preprint database (hep-th/9502100), a few interesting
papers \cite{BR95} appeared where the equivalence is shown
in the different method  by using the same interpolating
Lagrangian as derived in this paper.  However
they do not derive the interpolating Lagrangian and consider
it as a starting point without discussing its origin.

\end{document}